\documentclass[12pt]{article}
\usepackage{amsmath,amssymb}
\usepackage{graphicx,color}
\numberwithin{equation}{section}
\usepackage{cite}
\usepackage{bm}
\usepackage{dcolumn}  
\newcommand{\beq}{\begin{equation}}
\newcommand{\eeq}{\end{equation}}
\newcommand{\beqa}{\begin{eqnarray}}
\newcommand{\eeqa}{\end{eqnarray}}
\newcommand{\beqar}{\begin{eqnarray*}}
\newcommand{\eeqar}{\end{eqnarray*}}
\newcommand{\al}{\alpha}

\newcommand{\z}{\zeta}

\newcommand{\labell}[1]{\label{#1}} 

\begin{document}
\begin{titlepage}
\hfill
\vbox{
    \halign{#\hfil         \cr
           IPM/P-2008/042 \cr
           } 
      }  
\vspace*{20mm}
\begin{center}
{\Large {\bf On thermodynamics of ${\mathcal{N}}=6$ superconformal Chern-Simons theories at strong coupling}\\ }

\vspace*{15mm} 
\vspace*{1mm} 
{\large   Mohammad R. Garousi\footnote{garousi@mail.ipm.ir}, Ahmad Ghodsi\footnote{ahmad@mail.ipm.ir}  and Mehran Khosravi\footnote{me$\_$kh30@stu-mail.um.ac.ir}}
\vspace*{1cm}

%


{ {Department of Physics, Ferdowsi University, P.O. Box 1436, Mashhad, Iran}}\\
{ and}\\
{ Institute for Studies in Theoretical Physics and Mathematics (IPM)\\
P.O. Box 19395-5531, Tehran, Iran}\\
\vskip 2.6 cm

\end{center}

\begin{abstract} 
Recently it has been conjectured that ${\mathcal{N}}=6$, $U(N)_k\times U(N)_{-k}$ Chern-Simons theory is dual to M-theory on $AdS_4\times S^7/Z_{k}$.
By studying one-loop correction to the M-theory effective action, we calculate the correction to the entropy of thermal field theory at strong coupling. For large $k$ level, we have also found the $\alpha'$ correction to the entropy from the string correction of  the type IIA effective action. The structure of these two corrections at strong 't Hooft coupling are different. 
\end{abstract}

\end{titlepage} 

\section{Introduction}
Following the idea that the Chern-Simons gauge theory may be used to describe the coincident M2-branes \cite{Schwarz:2004yj}, Bagger and Lambert \cite{Bagger:2007vi} as well as Gustavsson \cite{Gustavsson:2007vu} have constructed three dimensional ${\mathcal{N}}=8$ superconformal $SO(4)$ Chern-Simons gauge theory based on 3-algebra.
It is believed that the BLG theory at level one describes two M2-branes on $R^8/Z_2$ orbifold \cite{Lambert:2008et}.

The signature of the metric on 3-algebra in the BLG model is positive definite. This assumption has been relaxed in \cite{Gomis:2008uv} to study $N$ coincident M2-branes with $N>2$. The so called BF membrane theory with arbitrary semi-simple Lie group has been proposed in \cite{Gomis:2008uv}. This theory has ghost fields, however, the unitarity returns by gauging the global shift symmetry for the Bosonic and Fermionic ghost fields \cite{Bandres:2008kj}. This gauged BF membrane theory is argued to be equivalent to the three dimensional ${\mathcal{N}}=8$ SYM theory \cite{Ezhuthachan:2008ch}.

Another approach to study multiple coincident M2-branes is proposed by Aharony-Bergman-Jafferies-Maldacena (ABJM) \cite{Aharony:2008ug}. They consider
a particular brane configuration which preserves  ${\mathcal{N}}=3$ supersymmetry. At low energy, by integrating out the massive modes of the brane configuration one finds $U(N)_k\times U(N)_{-k}$ Chern-Simons conformal field theory which preserves ${\mathcal{N}}=6$ supersymmetry. This theory is renormalizable and is consistent even at high energies. By lifting the brane configuration to M-theory they have shown that the gauge theory is equivalent to the low energy theory of $N$ coincident M2-branes in orbifold $R^8/Z_k$. Using the AdS-CFT correspondence, then they have proposed the duality between ${\mathcal{N}}=6$ Chern-Simons theory at level $k$ and M-theory on $AdS_4\times S^7/Z_k$.
For further study of the theory of multiple coincident M2-branes see \cite{M2}.

In this paper, we are interested in studying the entropy of the above gauge theory at finite temperature. At weak coupling, $\lambda\ll 1$, the entropy is given by the free field theory \cite{Aharony:2008ug}
\beq
S=3V_2N^2T^2(\frac{7\zeta(3)}{\pi}+O(\lambda))\,.
\eeq
At strong coupling and at small $k$ level, the entropy is given by the area of the Schwarzschild $AdS_4\times S^7/Z_k$ \cite{Aharony:2008ug}
\beq
S=V_2T^2\frac{2^\frac72}{3^3}\pi^2 N^2\frac{1}{\sqrt{\lambda}}\,.\label{entorb}
\eeq
The correction to the above  entropy in which we are interested, is coming from one loop corrections to the Schwarzschild $AdS_4\times S^7/Z_k$ solution. At strong coupling and at large $k$, one expects similar behavior for the entropy from the type IIA supergravity. We will also find (\ref{entorb}) and its $\alpha'$ correction. In the presence of higher derivative corrections, the entropy can be calculated using the Wald formula \cite{Wald:1993nt} or the free energy method \cite{Hawking:1982dh}.

An outline of the paper is as follows.
In section two we briefly review the p-brane solutions of M-theory at tree-level. In third section we compute the corrections to the Schwarzschild $AdS_4\times S^7$ in the presence of higher derivative corrections to the eleven dimensional supergravity action. In subsection (3.1) we calculate the entropy using the Wald formula and in subsection (3.2) we find the same result for entropy using the free energy method.
In section four, we find the entropy of the $U(N)_k\times U(N)_{-k}$ Chern-Simons conformal field theory at strong coupling and at small $k$, which is given by the the entropy of Schwarzschild $AdS_4\times S^7/Z_k$ at one-loop. In subsection (4.1),  we calculate the entropy of the thermal field theory at strong coupling and at large level $k$ from studying the type IIA supergravity, and find its $\alpha'$ correction. Finally in the last section, we calculate the entropy of Schwarzschild $AdS_7\times S^4$ at one-loop which is the entropy of the world-volume theory of $N$ coincident M5-branes at strong coupling.
\section{Review of M-theory}
The eleven dimensional supergravity action and its one loop corrections is given by 
\beq
S=\frac{1}{2\kappa_{11}^2}\int d^{11}x \sqrt{-g}\bigg(R-\frac{1}{2 n!}F_{(n)}^2+\gamma W\bigg)\,, \label{action}
\eeq
where $\gamma=4\pi^2\kappa_{11}^{\frac43}/3$ and $W$ in terms of the Weyl tensors is
\beq
W=C^{hmnk}C_{pmnq}{C_h}^{rsp}{C^q}_{rsk}+\frac12 C^{hkmn}C_{pqmn}{C_h}^{rsp}{C^q}_{rsk}\,. \label{W}
\eeq
One set of solutions to the tree-level part of the above action is 
the non-extremal p-brane solutions in $D=11$ dimensional space-time (see e.g. \cite{Petersen:1999zh})
\beqa
ds^2&=&H(r)^{-\frac{d-2}{D-2}}\bigg(-f(r)dt^2+\sum_{i=1}^{p}(dx^i)^2\bigg)
+H(r)^{\frac{p+1}{D-2}}\bigg(f(r)^{-1}dr^2+r^2d\Omega_{d-1}^2\bigg)\,, \cr &&\cr
F_{ti_1\cdots i_pr}&=&\epsilon_{ti_1\cdots i_pr}H(r)^{-2}\frac{Q}{r^{d-1}}\,,
\eeqa
where $D=(p+1)+d$ and $d$ is the number of dimensions transverse to the p-branes. The functions $H$ and $f$ are $H(r)=1+\left(\frac{L}{r}\right)^{d-2}$ and $f(r)=1-\left(\frac{r_0}{r}\right)^{d-2}$. In the above equation the relation between $L$ and $Q$ is 
\beqa
L^{2(d-2)}+L^{d-2}r_0^{d-2}&=&\frac{Q^2}{(d-2)^2}\,.\label{relation}
\eeqa
For $r_0=0$ we obtain the extremal solution, depending only on a
single parameter, $Q$, related to the charge
density of the BPS p-brane. For $r_0\neq 0$ a horizon develops at
$r=r_0$.
To obtain the near horizon geometry, we take the limit $r\ll
L$. In this limit the relation (\ref{relation}) simplifies to
$L^{d-2}=Q/(d-2)$, and the non-extremal solution becomes 
\beqa
ds^2\!\!\!&=&\!\!\!\left(\frac{r}{L}\right)^{\frac{2(d-2)}{p+1}}\!\!\left[
-\left(1-\left(\frac{r_0}{r}\right)^{d-2}\right)dt^2
+\sum_{i=1}^{p}(dx^i)^2\right]\!\! +\frac{\left(\frac{L}{r}\right)^{2}}{
\left(1-\left(\frac{r_0}{r}\right)^{d-2}\right)}dr^2
+L^2(d\Omega_{d-1})^2,\cr &&\cr
F_{ti_1\cdots i_pr}\!\!\!&=&\!\!\!(d-2)\epsilon_{ti_1\cdots
i_pr}\frac{r^{d-3}}{h^{d-2}}\,,\label{psol}
\eeqa 
which is the product space of $S^{d-1}$ with the Schwarzschild black hole in $AdS_{D-d+1}$. According to the AdS/CFT correspondence,
there is a finite temperature field theory dual to the M-theory on Schwarzschild black hole in $AdS_{D-d+1}\times S^{d-1}$. 
\section{M2-branes in flat space}
The field theory of large $N$ M2-branes in flat space is dual to M-theory on $AdS_4\times S^7$ and its finite temperature is dual to M-theory on Schwarzschild 
$AdS_4\times S^7$. At strong coupling, the filed theory is dual to eleven dimensional supergravity on Schwarzschild 
$AdS_4\times S^7$. The entropy and the temperature of field theory are given by the corresponding quantities in Schwarzschild 
$AdS_4\times S^7$, 
\beq
S= V_2T^2 \frac{2^\frac72}{3^3}\pi^2 N^\frac32\,,
\eeq
which is the area of the horizon of Schwarzschild 
$AdS_4\times S^7$. The background and subsequently the entropy and temperature of the field theory are modified by the one-loop corrections. 
By including the higher derivative terms one needs to 
consider a general ansatz for the metric and field strength, and then finds the deformations of the near horizon geometry of M2-branes.
We start with the following ansatz for the metric
\beq
ds^2=S^{2n}(r)H^2(r)\bigg(K^2(r)d\tau^2+P^2(r)dr^2+\sum_{i=1}^2dx_i^2\bigg)+L^2S^2(r)d\Omega_7^2\,, \label{ansatz}
\eeq
where $H(r)=\frac{r}{L}$. At tree-level, the above functions are
\beq
K(r)=(1-\frac{r_0^3}{r^3})^\frac12\,, \quad P(r)=\frac{L^2}{2r^2}(1-\frac{r_0^3}{r^3})^{-\frac12}\,,\quad S(r)=1\,,
\eeq
where we have made the space-time Euclidean and changed the variable $r$ in (\ref{psol}) to $r^2/L$. The electric field strength at tree-level is also given by $F_{\tau x_1x_2r}=3 \frac{r^2}{L^3}$.
Considering the fact that the charge of membranes is independent of the loop corrections, we restrict the field strength of M2-branes to the
following form
\beq
F_{\tau x_1x_2r}=\frac{6r^4}{L^5}K(r)P(r)S^{4n-7}(r)\,.\label{F4}
\eeq
Using the above ansatz, one writes the action (\ref{action}) as an integration over $r$
\beq
I=-\frac{\beta V_2V_7}{2\kappa_{11}^2}\int dr \bigg({\mathcal{L}}+\gamma {\mathcal{W}}\bigg)\,, 
\eeq
where 
\beqa
{\mathcal{L}}=S^{4n+7}{r^4}{L^3}KP\left(R-\frac12\frac{1}{4!}F_{(4)}^2\right)\,,\quad\quad
{\mathcal{W}}=S^{4n+7}{r^4}{L^3}KPW\,.
\eeqa
The Euler-Lagrange equations are
\beq
\frac{\partial {\mathcal{L}}}{\partial \Phi}-\frac{d}{dr}\frac{\partial {\mathcal{L}}}{\partial \Phi'}
+\frac{d^2}{dr^2}\frac{\partial {\mathcal{L}}}{\partial \Phi''}=-\gamma\bigg(\frac{\partial {\mathcal{W}}}{\partial \Phi}-\frac{d}{dr}\frac{\partial {\mathcal{W}}}{\partial \Phi'}+\frac{d^2}{dr^2}\frac{\partial {\mathcal{W}}}{\partial \Phi''}\bigg)\equiv -\gamma\omega_\Phi\,,\label{EL}
\eeq
where $(\Phi=\{K,P,S\})$.
The left hand side can be computed by inserting the ansatz (\ref{ansatz}) and (\ref{F4}) into the supergravity action. 
The right hand side of the Euler-Lagrange equation belong to the next order of perturbations so we just need to compute it by inserting the tree-level solutions. The value of $W$ is
\beq
W=\frac{1152}{L^8}\bigg(\frac{r_0^{12}}{r^{12}}-\frac{7}{50}\frac{r_0^6}{r^6}+\frac{1533}{4000}\bigg)\,.\label{WM2}
\eeq
In the above relation the first term is coming from $AdS_4$ and the other terms result from the fact that we have considered the eleven dimensional metric in which the radii of $AdS_4$ and $S^7$ are not equal. These terms makes the entropy to be different from the one that has been found in \cite{Gubser:1998nz} in which only the $AdS_4$ part has been considered.\footnote{Notice that in D3-brane case the radii of $AdS_5$ and $S^5$ are equal so that the value of $W$ for 5 dimension and 10 dimension are the same.}

The right hand sides for the Euler-Lagrange equations (\ref{EL}) are given by
\beqa
\omega_K\!\!\!&=&\!\!\!-\frac{6}{25}\frac{1}{L^3 r^{\frac{17}{2}}(r^3-r_0^3)^\frac12}\bigg(1533 r^{12} +3136 r_0^6 r^6 +129472 r^3 r_0^9 -160800 r_0^{12}\bigg)\,,\cr &&\cr
\omega_P\!\!\!&=&\!\!\!-\frac{12}{25}\frac{(r^3-r_0^3)^\frac12}{L^5 r^{\frac{19}{2}}}\bigg(1533 r^{12} -3136 r_0^6 r^6 +1344 r^3 r_0^9 -26400 r_0^{12}\bigg)\,, \\
\omega_S\!\!\!&=&\!\!\!-\frac{36792}{25}\frac{1}{L^3 r^{10}}\bigg(\!(n-\frac{29}{20}) r^{12} +\frac{224}{219}(n-\frac{47}{56}) r_0^6 r^6 -\frac{192}{73}(n-1) r^3 r_0^9 +\frac{800}{511}(n-\frac74) r_0^{12}\!\bigg).\nonumber
\eeqa
In order to solve the Euler-Lagrange equations we use the perturbative approach. We consider the solutions to be just some corrections to the tree-level solutions
\beqa
K(r)&=&(1-\frac{r_0^3}{r^3})^\frac12\left(1+\gamma X(r)\right)\,,\cr &&\cr
P(r)&=&\frac{L^2}{2r^2}(1-\frac{r_0^3}{r^3})^{-\frac12}\left(1+\gamma Y(r)\right)\,,\cr &&\cr 
S(r)&=&\left(1+\gamma Z(r)\right)\,.\label{perturb}
\eeqa
Inserting them into the left hand side of the Euler-Lagrange equations one finds the following differential equations for perturbed functions: 
\beqa
0\!\!\!\!&=&\!\!\!\!100L^6r^{10}(n+\frac72)\bigg[r(r^3-r_0^3)Z''(r)+3(r^3-\frac{r_0^3}{2})Z'(r)-3r^{2}Z(r)\bigg]-300L^6r^{12}Y(r)\cr &&\cr
\!\!\!\!&-&\!\!\!\!100L^6r^{10}(r^3-r_0^3)Y'(r)+4599r^{12}+9408r^6r_0^6+388416r^3r_0^9-482400r_0^{12}\,,\cr &&\cr
0\!\!\!\!&=&\!\!\!\!300L^6r^{10}(n+\frac72)\bigg[(r^3-\frac{r_0^3}{2})Z'(r)-r^{2}Z(r)\bigg]-300L^6r^{12}Y(r)\cr &&\cr
\!\!\!\!&+&\!\!\!\!100L^6r^{10}(r^3-r_0^3)X'(r)+4599r^{12}-9408r^6r_0^6+4032r^3r_0^9-79200r_0^{12}\,, \\
0\!\!\!\!&=&\!\!\!\!1500(n^2+7n+7)L^6r^{10}\bigg[r(r^3-r_0^3)Z''(r)+4(r^3-\frac{r_0^3}{4})Z'(r)-4\frac{n^2+7n-\frac{91}{8}}{n^2+7n+7}r^2Z(r)\bigg]\cr &&\cr
\!\!\!\!&+&\!\!\!\!500(n+\frac72)L^6r^{10}\bigg[r(r^3-r_0^3)X''(r)+5(r^3-\frac{r_0^3}{10})X'(r)-3(r^3-\frac{r_0^3}{2})Y'(r)-12r^2Y(r)\bigg]\cr &&\cr
\!\!\!\!&+&\!\!\!\!(91980n-133371)r^{12}+94080(n-\frac{47}{56})r^6r_0^6-241920(n-1)r^9r_0^3+144000(n-\frac74)r_0^{12}\,.\nonumber\label{diffM2}
\eeqa
There is an interesting solution to the above equations. If one considers 
\beq
n=-\frac72\,,
\eeq
the above equations simplify drastically, i.e. the $Z(r)$ dependence in the first two equations will be canceled. Moreover, for $n=-\frac72$ the area of the horizon in terms of $r_0$ is independent of $S(r)$. We shall show that this simplifies the Wald formula when calculating the entropy.  

To find the solution for above differential equations one should use the boundary condition at the horizon. Solving the first equation gives rise to the following exact value for $Y(r)$
\beq
Y(r)=\frac{1}{L^6}(\frac{1533}{100}+\frac{3568}{25}\frac{r_0^3}{r^3}+\frac{2784}{25}\frac{r_0^6}{r^6}-536\frac{r_0^9}{r^9})\,,
\eeq
where the constant of differential equation has been fixed by imposing the fact that the value of $Y(r)$ at horizon, $r=r_0$, must be finite. 
Inserting the value of $Y(r)$ into the second equation, one will find
\beq
X(r)=\frac{1}{L^6}(c-\frac{3568}{25}\frac{r_0^3}{r^3}-\frac{3568}{25}\frac{r_0^6}{r^6}+88\frac{r_0^9}{r^9})\,,
\eeq
where $c$ is a constant which can not be fixed by the boundary conditions. Instead, up to the first order of $\gamma$, it can be set to zero by time scaling.
Finally replacing all the above values in the third equation, one finds the following second order differential equation for $Z(r)$
\beqa
0&=&-7875L^6r^{11}(r^3-r_0^3)Z''(r)-31500L^6r^{10}(r^3-\frac14r_0^3)Z'(r)+141750L^6r^{12}Z(r)\cr &&\cr
&-&455301r^{12}-408240r^6r_0^6+1088640r^3r_0^9-756000r_0^{12}\,.\label{Z}
\eeqa
Unlike $X(r)$ and $Y(r)$ which have a simple series solution, here one can not find such a behavior for $Z(r)$. One may try to solve this differential equation with the well defined boundary conditions at the horizon and at infinity. The solution is needed for studying the Kaluza-Klein modes. However as we will see we do not need to know the exact form of $Z(r)$ for calculating the thermodynamical quantities such as temperature and entropy in which we are interested in this paper. 
\subsection{Entropy from the Wald formula}
One way of calculating the entropy in a higher derivative theory of gravity is the Wald formula \cite{Wald:1993nt}. It is given by  
\beq
S=\frac{4\pi}{2\kappa_{11}^2}\int dx^H\sqrt{g^H}\frac{\partial}{\partial R_{\mu\nu\rho\lambda}}(L+\gamma W)g^{\bot}_{\mu\rho}\,g^{\bot}_{\nu\lambda}\,,\label{Wald}
\eeq
where the superscript $H$ refers to the horizon and $g^{\bot}$ is the orthogonal metric to the horizon. In our model the orthogonal directions are $\tau$ and $r$, i.e. $g^{\bot}_{rr}=g_{rr}$ and $g^{\bot}_{\tau\tau}=g_{\tau\tau}$. As we have mentioned before for $n=-\frac 72$ the horizon area in terms of $r_0$ does not modify by the higher derivative corrections, so the first term in above formula which is the area of horizon at leading order does not modify either.
So we need only the tree-level solution to calculate the first term. On the other hand, the second term is proportional to $\gamma$, so to the first order of $\gamma$ one has to replace the tree-level solution into the second term too. Therefor the first term is 
\beq
S_1=\frac{4\pi}{2\kappa_{11}^2}V_2V_7L^5r_0^2\,,
\eeq
and the second term is 
\beq
S_2=\frac{4\pi}{2\kappa_{11}^2}V_2V_7L^5r_0^2(\frac{34616}{250}\frac{\gamma}{L^6})\,.
\eeq
The final result for entropy will be
\beq
S=\frac{4\pi}{2\kappa_{11}^2}V_2V_7L^5r_0^2(1+\frac{34616}{250}\frac{\gamma}{L^6})\,.
\eeq
To write the above entropy in terms of the temperature, we recall that the temperature of the black hole is given by the surface gravity at the horizon 
\beq
\hat\kappa=2\pi T=\sqrt{g^{rr}}\frac{d}{dr}\sqrt{g_{\tau\tau}}|_{r=r_0}=\frac{1}{rP}\frac{d}{dr}(rK)+\frac{nK}{PS}\frac{d}{dr}S|_{r=r_0}\,.
\eeq
As we have anticipated before the temperature is independent of $Z(r)$ because the second term is zero. Note that at horizon $\frac{dS}{dr}=\gamma\frac{dZ}{dr}$ is finite (\ref{Z}) and $K/P=0$. However the temperature does depend on corrections of $K$ and $P$. So the entropy in terms of temperature depends on the loop correction of $K$ and $P$ and is independent of $Z$. The temperature is 
\beq
T=\frac{3}{2\pi}\frac{r_0}{L^2}(1+\frac{\gamma}{L^6}\frac{1383}{20})\,,
\eeq
and the entropy in terms of temperature is 
\beq
S=\frac{V_2V_7}{2\kappa_{11}^2}(\frac{16}{9}\pi^3L^9T^2)(1+\frac{41}{250}\frac{\gamma}{L^6})\,,\label{entr0}
\eeq
Using the relations $L^9=N^\frac32 \kappa_{11}^{2}\frac{\sqrt{2}}{\pi^5}$, where $N$ is the total number of membranes and $V_7=\frac{\pi^4}{3}$
one finds
\beq
S=V_2T^2\bigg\{\frac{2^{\frac72}}{3^3}\pi^2N^\frac32+\frac{41}{250}\left(\frac{2\pi}{3}\right)^5 2^\frac16 3 \pi^\frac73 N^\frac12\bigg\}\,.
\eeq
Note that the coefficient of the last term, $\frac{41}{250}$, is different from the value found in \cite{Gubser:1998nz}. This is resulted from the fact that we have considered eleven dimensional metric in which $W$ is given by (\ref{WM2}) instead of four dimensional metric considered in \cite{Gubser:1998nz} in which $W$ is given by the first term of (\ref{WM2}). To double check our results we calculate the entropy from the free energy in the next section.
\subsection{Entropy from free energy}
Another way of finding the entropy of a higher derivative theory is to use the free energy. Following \cite{Hawking:1982dh}, one can identify the free energy of the theory with the Euclidean gravitational action, $I$, times the temperature, $T$, i.e.
\beq
I=\beta F\,,
\eeq  
where $\beta=\frac{1}{T}$. 
The calculation of the Euclidean action is divergent at large distances,$r_{max}$, and requires a subtraction. The integral must be regulated by subtracting off
its zero temperature limit, i.e.
\beq
F=\lim_{r_{max}\rightarrow\infty}\frac{I-I_{0}}{\beta}\,,
\eeq
where $I_0$ is the zero temperature limit in which the periodicity of the Euclidean time is defined by $\beta'$. One must adjust $\beta'$ so that
the geometry at $r=r_{max}$ is the same in the two cases, i.e. the black hole and its zero temperature. This can be done by equating the circumference of the Euclidean time in two cases 
\beq
\int_0^{\beta'}\sqrt{g^{{\rm (zero\, temperature)}}_{\tau\tau}}|_{r_{max}}d\tau=\int_0^\beta\sqrt{g^{{\rm (black\, hole)}}_{\tau\tau}}|_{r_{max}}d\tau.
\eeq
Having found $F$, the entropy in terms of the free energy is then given by $S=-\frac{\partial F}{\partial T}$.

Let us now try to find the entropy of membranes with this approach. To simplify the calculations we use the observation made in the previous section that the entropy and temperature are independent of $Z(r)$. So we set $Z(r)$ equal to zero in this approach. The Euclidean action and its regulator are 
\beqa
I\!\!\!&=&\!\!\!-\frac{\beta V_2V_7}{2\kappa_{11}^2}\int_{r_0}^{r_{max}}dr({\mathcal{L}+\gamma{\mathcal{W}}})\cr &&\cr
\!\!\!&=&\!\!\!4\frac{\beta V_2V_7}{2\kappa_{11}^2} L^{3}(r_{max}^3-r_0^3)\bigg[1-\frac{\gamma}{L^6}(\frac{32193}{500}+\frac{6968}{25}\frac{r_0^3}{r_{max}^3}+\frac{4784}{25}\frac{r_0^6}{r_{max}^6}+48\frac{r_0^9}{r_{max}^9})\bigg]\,,\cr &&\cr
I_0\!\!\!&=&\!\!\!-\frac{\beta' V_2V_7}{2\kappa_{11}^2}\int_{0}^{r_{max}}dr({\mathcal{L}+\gamma{\mathcal{W}}})\cr &&\cr
\!\!\!&=&\!\!\!4\frac{\beta' V_2V_7}{2\kappa_{11}^2}L^{3}r_{max}^3\bigg[1-\frac{\gamma}{L^6}(\frac{32193}{500})\bigg]\,,
\eeqa 
and the relation between $\beta$ and $\beta'$ is given by $\beta'=\beta K(r)|_{r_{max}}$.
The free energy then becomes
\beqa
F=\lim_{r_{max}\rightarrow\infty}\frac{I-I_{0}}{\beta}=-\frac{V_2V_7}{2\kappa_{11}^2}2r_0^3L^3(1+\frac{103807}{500}\frac{\gamma}{L^6})\,,
\eeqa
In terms of temperature it becomes
\beq
F=-\frac{V_2V_7}{2\kappa_{11}^2}(\frac{16}{27}\pi^3L^9T^3)(1+\frac{41}{250}\frac{\gamma}{L^6})\,.\label{freem2T}
\eeq
Finally, the entropy is given by 
\beq
S=3\frac{V_2V_7}{2\kappa_{11}^2}(\frac{16}{27}\pi^3L^9T^2)(1+\frac{41}{250}\frac{\gamma}{L^6})\,.
\eeq
This is exactly the entropy that we have found in (\ref{entr0}) using the Wald formula. 
\section{M2 branes in orbifold space}
It has been conjectured that ${\mathcal{N}}=6$ superconformal Chern-Simons matter theories at level $k$ with gauge group $U(N)\times U(N)$ is dual to M-theory on $AdS_4\times S^7/Z_k$ \cite{Aharony:2008ug}.
At finite temperature and at strong 't Hooft coupling $\lambda=N/k$, the entropy of field theory is given by the entropy of Schwarzschild black hole in $AdS_4\times S^7/Z_k$, (\ref{entorb}). 
In this section we are going to find the correction to this entropy from higher derivative corrections in the gravity side. In the presence of higher derivative terms, the Schwarzschild $AdS_4\times S^7/Z_k$ background is changing. We consider the same ansatz for the $AdS_4$ part as before. 
So we choose the following ansatz for the eleven dimensional metric
\beq
ds^2=S^{-7}(r)H^2(r)\bigg(K^2(r)d\tau^2+P^2(r)dr^2+\sum_{i=1}^2dx_i^2\bigg)+L^2S^2(r)ds_{S^7/Z_k}^2\,, \label{ansatzorb}
\eeq
where the orbifold $S^7/Z_k$ can be written as follows (see e.g. \cite{orb1} or \cite{orb2})
\beq
ds_{S^7/Z_k}^2=ds^2_{CP^3}+(d\alpha+\omega)^2,
\eeq
where
\beqa
\omega&=&\frac{1}{2}\bigg(\cos^2\xi-\sin^2\xi\bigg)d\psi +\frac{1}{2}\cos^2\xi
\cos\theta_1 d\phi_1+ \frac{1}{2}\sin^2\xi \cos\theta_2 d\phi_2 \,,\cr &&\cr
ds^2_{CP^3}&=&d\xi^2+\cos\xi^2\sin^2\xi\bigg(d\psi+\frac{\cos\theta_1}{2}d\phi_1-
\frac{\cos\theta_2}{2}d\phi_2\bigg)^2\cr &&\cr
&+&\frac{1}{4}\cos^2\xi\bigg(d\theta_1^2+\sin^2\theta_1
d\phi_1^2\bigg)+\frac{1}{4}\sin^2\xi\bigg(d\theta_2^2+\sin^2\theta_2
d\phi_2^2\bigg)\,.
\eeqa
and the variables run the values $0\leq
\xi <\frac{\pi}{2}$, $0\leq \chi_i <4\pi$, $0\leq \phi_i \leq 2\pi$ and
$0\leq \theta_i<\pi$. The $Z_k$ orbifold in $\alpha$ direction gives a periodicity as $\alpha\sim \alpha+\frac{2\pi}{k}$.
The metric of $S^7/Z_k$ in terms of the above variables simplifies to
\beqa
ds^2_{S^7/Z_k}&=&d\alpha^2+d\xi^2+\frac{1}{4}(d\psi^2+\cos^2\xi d\phi_1^2+\sin^2\xi d\phi_2^2+\cos^2\xi d\theta_1^2+\sin^2\xi d\theta_2^2)\cr &&\cr
&+&\frac12(\cos^2\xi-\sin^2\xi)d\alpha d\psi+\frac12\cos^2\xi\cos\theta_1 d\alpha d\phi_1+\frac12\sin^2\xi\cos\theta_2 d\alpha d\phi_2 \cr &&\cr
&+&\frac14\cos^2\xi\cos\theta_1 d\psi d\phi_1-\frac14\sin^2\xi\cos\theta_2 d\psi d\phi_2\,.
\eeqa 
Note that the volume of $S_7/Z_k$ in terms of the volume of $CP_3$ is Vol$(S_7/Z_k)=\frac{2\pi}{k}$ Vol$(CP_3)$. With the above ansatz the differential equations in (\ref{diffM2}) and the temperature do not change. However the free energy (\ref{freem2T}) changes by a factor of $\frac{1}{k}$ due to integration over $\alpha$, i.e. 
\beq
F=-\frac{1}{k}\frac{V_2V_7}{2\kappa_{11}^2}(\frac{16}{27}\pi^3L^9T^3)(1+\frac{41}{250}\frac{\gamma}{L^6})\,.
\eeq
To write the free energy in terms of the number of M2-branes, we note that the total number of M2-branes is given by 
\beq
N=\frac{1}{2\kappa_{11}^2}\int_{S^7/Z_k}*F_{(4)}=\frac{1}{k}\frac{1}{2\kappa_{11}^2}\int_{S^7}*F_{(4)}\,,
\eeq
this gives $L^9=(kN)^\frac32 \kappa_{11}^{2}\frac{\sqrt{2}}{\pi^5}$. So the free energy becomes
\beq
F=-\frac{1}{k}V_2T^3\bigg\{\frac{2^{\frac72}}{3^4}\pi^2(kN)^\frac32+\frac{41}{250}\left(\frac{2\pi}{3}\right)^5 2^\frac16  \pi^\frac73 (kN)^\frac12\bigg\}\,.
\eeq
In terms of the 't Hooft coupling it becomes
\beq
F=-V_2T^3\bigg\{\frac{2^{\frac72}}{3^4}\pi^2\frac{N^2}{\sqrt{\lambda}}+\frac{41}{250}\left(\frac{2\pi}{3}\right)^5 2^\frac16  \pi^\frac73 \sqrt{\lambda}\bigg\}\,.\label{freeorb}
\eeq
The first term is the same entropy as in (\ref{entorb}), and the second term is the one loop correction in the gravity side. Note that this expression is valid for strong 't Hooft coupling and for large $N$, i.e. the second term is small with comparing to the first term.

\subsection{Large $k$ limit}
The radius of $\alpha$ circle is large whenever $k^5\ll N$, so the M-theory description is valid for small $k$.
However for larger value of $k$ the circle becomes small, the curvature becomes large, and so all higher derivative terms of M-theory are relevant to our calculation of entropy. Hence,  in this case, one should reduce the M-theory to a weakly coupled
IIA string theory. So consider the reduction of the M-theory background to the string frame of IIA as 
\beq
ds^2_{D=11}=e^{-2\phi/3}ds^2_{IIA}+e^{4\phi/3}(R_{11}d\tilde{\alpha}+A)^2\,,
\eeq
where $R_{11}=g_s^{2/3} l_p$ is the radius of the eleventh direction with angle $\tilde{\alpha}$. Comparing the above metric with (\ref{ansatzorb}), one finds  $\tilde{\alpha}=k\alpha$, the string coupling to be  $g_s^2e^{2\phi}=g_s^2L^3/(R_{11}^3k^3)=2^{5/2}\pi\sqrt{N/k^5}$, and the metric in IIA string theory  to be $(kR_{11}/L)ds^2_{IIA}=ds_{AdS_4}^2+L^2ds^2_{CP^3}$. 
Reduction also gives a two form magnetic field strength as
\beq
F_{(2)}=  dA=k d\omega\,.
\eeq
Additionally, we have a four form electric field strength as before which is $F_{\tau x_1 x_2 r}=\frac{3r^2}{L^3}$.

By looking at the metric one finds that the scalar curvatures behave as ${\mathcal{R}}_{AdS}\sim {\mathcal{R}}_{CP^3}\sim kR_{11}/L^3\sim \sqrt{k/N}$.
So if $k^5\gg N$ and at the same time $k\ll N$, one can ignore the string loop corrections and the higher derivative terms resulting from the $\alpha'$ corrections. The first $\alpha'$ correction to the type IIA effective action can be written in terms of $W$. 
Considering  these higher derivative corrections, we are going to find  their effects on the background solution and subsequently on the entropy.   So consider 
the Lagrangian of type II theory in the presence of first $\alpha'$ correction which is given by
\beqa 
S=-\frac{1}{16\pi G_{10}}\int d^{10}x\,\sqrt{g}\bigg\{e^{-2\phi}(R+4(\partial\phi)^2)-\frac12\frac{1}{4!}F_{(4)}^2-\frac12\frac{1}{2!}F_{(2)}^2
+\gamma e^{-2\phi}W\bigg\}\,,\labell{eff1}
\eeqa 
where $\gamma=\frac18\zeta(3)(\al')^3$.  We choose the following ansatz for the metric
\beq
ds_{IIA}^2=\left(\frac{L}{kR_{11}}\right)\left(S^{-6}(r)(\frac{r}{L})^2\bigg(K^2(r)d\tau^2+P^2(r)dr^2+\sum_{i=1}^2dx_i^2\bigg)+L^2S^2(r)ds_{CP3}^2 \right)\,,
\eeq
and for the electric four form 
\beq
F_{\tau x_1 x_2 r}=\frac{6r^4}{L^5}K(r)P(r)S^{-18}(r)\,.
\eeq
Again using the perturbative method (\ref{perturb}) we are able to compute the corrections to the supergravity solution, i.e.,
\beqa
K(r)&=&(1-\frac{r_0^3}{r^3})^\frac12\bigg[1+\frac{\gamma}{L^6}\left(\frac{kR_{11}}{L}\right)^3(-136\frac{r_0^3}{r^3}-136\frac{r_0^6}{r^6}+88\frac{r_0^9}{r^9})\bigg]\,,\cr &&\cr
P(r)&=&\frac{L^2}{2r^2}(1-\frac{r_0^3}{r^3})^{-\frac12}\bigg[1+\frac{\gamma}{L^6}\left(\frac{kR_{11}}{L}\right)^3(4+136\frac{r_0^3}{r^3}+136\frac{r_0^6}{r^6}-536\frac{r_0^9}{r^9})\bigg]\,,
\eeqa
and the differential equation for $Z(r)$  is given by
\beq
0=-3r^{11}L^6(r^3-r_0^3)Z''(r)-3L^6r^{10}(r^3-r_0^3)Z'(r)-420r^{12}L^6Z(r)+816r^{12}+2592r_0^{12}\,.
\eeq
The correction of the dilaton field does not enter into the entropy in which we are interested.

We are going to calculate the entropy from the Wald formula (\ref{Wald}). In terms of $r_0$ it is
\beq
S=\frac{4\pi V_2V_{CP^3}L^5r_0^2}{2kR_{11}\kappa_{10}^2}\left(1+112\frac{\gamma}{L^6}\left(\frac{kR_{11}}{L}\right)^3\right)\,,
\eeq  
where the first term is the area of the horizon and the second term is coming from the $W$ terms. The temperature in terms of $r_0$ is
\beq
T=\frac{3r_0}{2\pi L^2}\left(1+76\frac{\gamma}{L^6}\left(\frac{kR_{11}}{L}\right)^3\right)\,,
\eeq
and the entropy in terms of temperature becomes
\beq
S=\frac{8\pi^3L^9V_2V_{CP^3}T^2}{9kR_{11}\kappa_{10}^2}\left(1-40\frac{\gamma}{L^6}\left(\frac{kR_{11}}{L}\right)^3\right)\,.
\eeq
Using the relations $\frac{2\pi R_{11}}{ \kappa_{11}^2}=\frac{1}{\kappa_{10}^2}$,  $2\pi$ Vol$(CP^3)=$ Vol$(S^7)$, $l_{11}=g_2^{1/3} l_s$ and  $2\kappa^2_{11}=(2\pi l_{11})^9/2\pi$    one finds
\beq
S=\frac{2^{\frac72}\pi^2}{3^3}\frac{N^2}{\sqrt{\lambda}}V_2T^2\left\{1-\frac{5\z(3)}{\pi^{3}2^{\frac{17}{2}}\lambda^{3/2}}\right\}\,.
\eeq
Here also the result is valid for the strong 't Hooft coupling and for large $N$. The first term is the same as the corresponding term in the M-theory.  The second term is resulted from the first $\alpha'$ correction to the type IIA supergravity action. The structure of this term is very different from the one loop correction in the M-theory side. 
\section{M5-branes in flat space}
The thermal field theory of $N$ M5 branes in flat space is dual to M-theory on Schwarzschild $AdS_7\times S^4$. The entropy of the field theory at the strong coupling is given by the entropy of the tree-level Schwarzschild $AdS$ solution,
\beq
S=V_5T^5\frac{2^7}{3^6}\pi^3 N^3\,.
\eeq 
In this section we would like to find the one loop corrections to this entropy.
The near horizon geometry of N M5-branes at the supergravity level is $AdS_7$ Schwarzschild times $S^4$. 
By including the higher derivative terms we need to 
consider a general ansatz and again find deformations of this near horizon geometry. Using the observation in M2-brane case that there is an ansatz 
in which the area of horizon in terms of the parameter $r_0$ is independent of the higher derivative corrections, we use the following ansatz for M5- branes
\beq
ds^2=S^{-\frac85}(r)H^2(r)\bigg(K^2(r)d\tau^2+P^2(r)dr^2+\sum_{i=1}^5dx_i^2\bigg)+L^2S^2(r)d\Omega_4^2\,, \label{ansatzM5}
\eeq
where $H(r)=\frac{r}{L}$. At tree-level the above functions are
\beq
K(r)=(1-\frac{r_0^6}{r^6})^\frac12\,, \quad P(r)=\frac{2L^2}{r^2}(1-\frac{r_0^6}{r^6})^{-\frac12}\,,\quad S(r)=1\,,
\eeq
where we have changed the variable $r$ in (\ref{psol}) to $\sqrt{rL}$ and made the space-time Euclidean.
The electric field strength at tree-level is also given by $F_{\tau x_1...x_5r}=6 \frac{r^5}{L^6}$.
Considering the fact that the charge of M5-branes is independent of the loop corrections, we restrict the field strength of M5-branes to the
following form
\beq
F_{\tau x_1...x_5r}=\frac{3r^7}{L^8}K(r)P(r)S^{\frac{51}{5}}\,.
\eeq
The Euclidean action then simplifies to
\beq
I=-\frac{\beta V_5V_4}{2\kappa_{11}^2}\int dr \bigg({\mathcal{L}}+\gamma {\mathcal{W}}\bigg) \,,
\eeq
where 
\beqa
{\mathcal{L}}=S^{-\frac85}{r^7}{L^{-3}}KP\bigg(R-\frac12\frac{1}{7!}F_{(7)}^2\bigg)\,,\quad\quad
{\mathcal{W}}=S^{-\frac85}{r^7}{L^{-3}}KPW\,.
\eeqa
Here also the right hand side of the Euler-Lagrange equation belong to the next order of perturbations so we just need to compute it by inserting the tree-level solutions. The value of $W$ is
\beq
W=\frac{9}{8000L^8}\bigg(1533+7000\frac{r_0^{12}}{r^{12}}-11000\frac{r_0^{18}}{r^{18}}+45625\frac{r_0^{24}}{r^{24}}\bigg)\,.
\eeq
Note that only the last term is the value of $W$ for seven dimensional Schwarzschild $AdS_7$ considered in \cite{Gubser:1998nz}.
The derivatives of ${\mathcal{W}}$ in (\ref{EL}) which appear as source terms for the equations of motion are
\beqa
\omega_K&=&\frac{3}{4000}\frac{1}{L^9 r^{16}(r^6-r_0^6)^\frac12}\bigg(6298875 r_0^{24} -5507000 r_0^{18} r^6 +121400 r_0^{12} r^{12} +6351 r^{24}\bigg)\,,\cr &&\cr
\omega_P&=&\frac{3}{8000}\frac{(r^6-r_0^6)^\frac12}{L^{11} r^{20}}\bigg(1030875 r_0^{24} -181000 r_0^{18} r^6 +63400 r_0^{12} r^{12} +6351 r^{24}\bigg)\,, \cr &&\cr
\omega_S&=&-\frac{27}{2500}\frac{1}{L^9 r^{19}}\bigg(4599 r^{24}-44000 r^{12} r_0^{12}+234000 r^6 r_0^{18}-243625 r_0^{24}\bigg)\,.
\eeqa
Using the perturbative method to solve the  Euler-Lagrange equations, i.e.
\beqa
K(r)&=&(1-\frac{r_0^6}{r^6})^\frac12\left(1+\gamma X(r)\right)\,,\cr &&\cr
P(r)&=&\frac{2L^2}{r^2}(1-\frac{r_0^6}{r^6})^{-\frac12}\left(1+\gamma Y(r)\right)\,,\cr &&\cr 
S(r)&=&\left(1+\gamma Z(r)\right)\,,
\eeqa
one finds the following differential equations 
\beqa
0&=&-20000 r^{19}(r^6-r_0^6) L^6 Y'(r)-120000 r^{24} L^6 Y(r) \cr &&\cr
&-&19053 r^{24}-364200 r^{12} r_0^{12}+16521000 r^6 r_0^{18}-18896625 r_0^{24}\,,\cr &&\cr
0&=&20000 r^{19} L^6(r^6-r_0^6)^2 X'(r)-120000r^{24}L^6(r^6-r_0^6)Y(r)\cr &&\cr
&-&19053 r^{30}+19053 r^{24} r_0^6+733200 r^{12} r_0^{18}-3635625 r^6 r_0^{24}-190200 r^{18} r_0^{12}+3092625 r_0^{30}\,,\cr &&\cr
0&=&-28800 r^{19}L^6\bigg(r(r^6-r_0^6)Z''(r)+(7r^6-r_0^6)Z'(r)-72 r^5Z(r)\bigg)\cr &&\cr
&+& \frac{993384}{5} r^{24}-10524600 r_0^{24}+10108800 r^6 r_0^{18}-1900800 r^{12} r_0^{12}\,.
\eeqa
The solution to the first two equations are
\beqa
X(r)&=&-\frac{1}{L^6}(c+\frac{21299}{1600}\frac{r_0^6}{r^6}+\frac{4999}{320}\frac{r_0^{12}}{r^{12}}-\frac{2749}{320}\frac{r_0^{18}}{r^{18}})\,,\cr &&\cr
Y(r)&=&-\frac{1}{L^6}(\frac{6351}{40000}-\frac{21299}{1600}\frac{r_0^6}{r^6}-\frac{5231}{320}\frac{r_0^{12}}{r^{12}}+\frac{16797}{320}\frac{r_0^{18}}{r^{18}})\,,
\eeqa
where $c$ is a constant which can be set to zero by time scaling, up to first order of $\gamma$. Here also the differential equation for $Z(r)$ does not have such a simple solution. However we do not need to know this solution for finding the thermodynamical quantities $T$ and $S$ in which we are interested. So we ignore $Z(r)$ from now on in this section.

The temperature of the black hole can be found by surface gravity at the horizon 
\beq
\hat\kappa=2\pi T=\sqrt{g^{rr}}\frac{d}{dr}\sqrt{g_{\tau\tau}}|_H=\frac32\frac{r_0}{L^2}(1+\frac{\gamma}{L^6}\frac{105901}{40000})\,.
\eeq
Using the ansatz (\ref{ansatzM5}), one observes that the horizon area in terms of $r_0$ does not modify by the higher derivative corrections so the first term in the Wald formula (\ref{Wald}), which is the area of horizon at leading order does not modify either.
So we need only the tree-level solution to calculate the first term in the Wald formula (\ref{Wald}) 
\beq
S_1=\frac{4\pi}{2\kappa_{11}^2}V_5V_4\frac{r_0^5}{L}\,.
\eeq
On the other hand the second term is proportional to $\gamma$, so to first order of $\gamma$ one has to replace the tree-level solution into the second term too, i.e.
\beq
S_2=\frac{4\pi}{2\kappa_{11}^2}V_5V_4\frac{r_0^5}{L}(\frac{12597}{500}\frac{\gamma}{L^6})\,.
\eeq
So the final result for the entropy in terms of the temperature is
\beq
S
=\frac{V_5V_4}{2\kappa_{11}^2}(\frac{2^{12}}{3^5}\pi^6L^9T^5)(1+\frac{95651}{8000}\frac{\gamma}{L^6})\,.\label{entm5}
\eeq
Inserting $L^9=N^3 \kappa_{11}^2 2^{-7}\pi^{-5}$ where $N$ is the total number of M5-branes and $V_4=8\frac{\pi^2}{3}$, one finds
\beq
S=V_5T^5\bigg(\frac{2^7}{3^6}\pi^3N^3+\frac{95651}{8000}\left(\frac{2\pi}{3}\right)^8 3\left(\frac{\pi}{2}\right)^\frac13 2^6 N \bigg)\,.
\eeq
Using the same steps as before for obtaining the free energy, one finds the following result for the free energy
\beqa
F
&=&-\frac{V_5V_4}{2\kappa_{11}^2}\frac12 \frac{r_0^6}{L^3}(1+\frac{1113661}{40000}\frac{\gamma}{L^6})\cr &&\cr
&=&-\frac{V_5V_4}{2\kappa_{11}^2}(\frac{2048}{729}\pi^6L^9T^6)(1+\frac{95651}{8000}\frac{\gamma}{L^6})\,,
\eeqa
which gives exactly the same entropy as in (\ref{entm5})

\end{document}